\documentstyle[twocolumn,aps,epsfig]{revtex}
\begin{document}
\draft
\title{Landau Expansion for the Critical Point of a Polydisperse System}

\author{C.\ Rasc\'{o}n$^{\star\,\dagger}$ and M.\ E.\ Cates$^{\star}$}
\address{
$^\star$ School of Physics, University of Edinburgh,\\
Mayfield Road, Edinburgh EH9 3JZ, United Kingdom\\
$^\dagger$ GISC. Departamento de Matem\'{a}ticas,
Universidad Carlos III de Madrid. \\ Avenida de la Universidad 30.
28911 Legan\'{e}s, Madrid. Spain.
}
\maketitle

\begin{abstract}
The effect of polydispersity on the phase diagram of a simple binary mixture 
is to split the binodal curve into cloud and shadow curves that cross at the 
critical point (which, in general, is not at the maximum of either curve). 
Recent theories 
of polydispersity have shown, in favorable cases, how to project the 
(infinite-dimensional) free energy of the polydisperse system into a low 
dimensional space of `moment densities'. We address here the issue of how to 
construct a Landau expansion from the projected free energy. For the simplest 
case where the excess free energy depends on one moment density $\rho_1$ (this 
includes Flory Huggins 
theory for length-polydisperse chains) we show that the minimal expansion 
remains quartic in $\rho_1$ but nonetheless has seven independent coefficients, 
not two. When this expansion is handled correctly all the peculiar qualitative 
features of 
the polydisperse phase diagram are recovered, as are the slopes of the cloud 
and shadow curves, and the curvature of the cloud. However, to 
correctly predict the curvature of the shadow, certain fifth order terms must 
be retained. We also consider the phase 
diagram on the temperature-pressure plane, where the coexistence line broadens 
into a region. In general, the critical point lies between the point of maximum 
temperature and the point of maximum pressure on the perimeter of this region. 
This behavior is again captured by the Landau expansion. 
\end{abstract}
\pacs{ PACS numbers: }

\section{Introduction}
Many systems, such as colloidal suspensions, and polymer blends and solutions, 
contain particles with an effectively continuous distributions of an attribute 
($\sigma$, say) such as size, charge, or chemical character. Such systems are 
termed polydisperse \cite{Review}. Their phase equilibria can become 
complicated, but may exhibit relatively simple features including coexistence 
of two liquid phases of different composition. In some cases the tendency to 
phase separate is inherited from the underlying monodisperse physics (e.g. 
attractive colloids); in others it stems only from polydispersity (e.g. 
fractionation among chains with a spectrum of chemical compositions). 

In a polydisperse system undergoing fluid-fluid demixing, the phase diagram 
looks rather different from that of a simple binary mixture even if the 
underlying physics is similar. It is common experimentally to draw the phase 
diagram on a composition/temperature plane; one effect of polydispersity is 
then to split the binodal curve into cloud and shadow curves that cross at the 
critical point. (The latter is not at the maximum of either curve.) This is 
because, when a system is cooled to the point where it becomes globally unstable 
to phase separation (the cloud point), the incipient new phase (its shadow) 
differs from the old one not only in the number of particles per unit volume, 
as usual, but in its $\sigma$-distribution. Because of this difference, the 
incipient new phase does not itself represent a point on the cloud curve. 
(However, it does lie on the cloud curve of a {\it different system} whose 
overall size distribution coincides with that of the incipient, shadow phase).

An interesting question is whether all this behavior can be captured within a 
Landau expansion. There are at least two reasons to consider this. 
Firstly, conceptual (as well as practical) progress has been made recently in 
understanding phase equilibria in polydisperse systems within a certain class 
of mean-field models: those in which the excess free energy depends on the 
$\sigma$ distribution only through a finite set of moments \cite{Review}. 
A promising method involves projecting the free energy onto a relatively small 
number of densities; when these are chosen correctly, the cloud, shadow, and 
spinodal curves, and with them any critical points, are found without 
approximation beyond those used to set up the mean-field free energy in the 
first place. 

Due to their mean-field character and restricted set of relevant densities, 
such projection techniques should be reconcilable with a Landau expansion near 
the critical point -- yet the unusual topology of the phase diagram appears, 
at first sight, to preclude this. In particular, the usual adoption of a 
quartic expansion in one density (with coefficients dependent only on 
temperature) is bound to exhibit a critical point at the apex of a binodal 
curve. We show below that a more careful Landau expansion does not suffer from 
this restriction, and instead reproduces the correct, more complicated 
behavior outlined above.

A second reason to study the Landau expansion is in the hope of going beyond 
mean field theory. Experimental values for the critical exponents of a
polydisperse system \cite{Experiment} deviate from those of the monodisperse
case and are consistent with Fisher-renormalised exponents \cite{FisherRen}. 
However, to our knowledge, there is no explicit renormalization group 
calculation for a polydisperse system. Also, quite apart from 
calculating critical indices, it is important for many purposes (see e.g. 
\cite{WildingFields}) to properly classify and identify the scaling fields near 
the critical point, and this task cannot really be begun without a consistent 
description at the Landau level. Hence the Landau expansion presented below 
represents a first step towards a fuller understanding of polydisperse critical 
behavior beyond mean field level.

\section{Monodisperse system}

In this section we review briefly the classical Landau 
expansion of the free energy around the critical point
for a simple monodisperse system. This will facilitate a
comparison with the polydisperse case and establish the
notation. 

In a canonical ensemble, the free energy $\cal F$ 
is defined as a function of the temperature $T$, the number of 
particles $N$ and the volume $V$. In the thermodynamic limit,
however, we can define a reduced free energy (density) as
\begin{eqnarray}
F(\rho,T)=\frac{{\cal F}(N,T,V)}{V\,k_{B}T},
\end{eqnarray}
which is a function only of the density of particles $\rho\!=\!N/V$
and the temperature. At fixed temperature, any terms that are constant or 
linear in density are irrelevant for obtaining the phase
equilibrium of the system (which is our main aim) since they can
be assimilated into shifts of the entropy and of the chemical potential. 

\subsection{Critical Point}
 
Within the mean field approximation, which is where the 
Landau expansion holds, the critical point must obey 
one condition common to all points on the spinodal curve,
\begin{eqnarray}
\label{MSpinodal}
\left.\frac{\partial^ 2F}{\partial\rho^2}\right|_{c} = 0,
\end{eqnarray}
and a second condition particular to the critical point,
\begin{eqnarray}
\label{MCritPoint}
\left.\frac{\partial^ 3F}{\partial\rho^3}\right|_{c} = 0,
\end{eqnarray}
where $|_c$ means evaluated at the critical point $(\rho^{\,c},T_c)$.

\subsection{Landau Expansion}

Once the critical point is determined, the expansion follows as usual:

\begin{eqnarray}
\label{MEnergy1}
F(\rho,T) \approx 
\frac{1}{2}\,\left.\frac{\partial^ 2F}{\partial\rho^2}
\right|_{\rho^c,T}{\delta\rho}^2 + \phantom{\hspace{1cm}} \nonumber \\
\frac{1}{3!}\,\left.\frac{\partial^ 3F}{\partial\rho^3}
\right|_{\rho^c,T}{\delta\rho}^3 + 
\frac{1}{4!}\,\left.\frac{\partial^ 4F}{\partial\rho^4}
\right|_{\rho^c,T}{\delta\rho}^4 + \dots,
\end{eqnarray}
where $\delta\rho\equiv\,\rho-\rho^c$. We forget about the
constant and linear terms since, as described above, they can be viewed as
(temperature dependent) shifts in the entropy and chemical
potential with no effect on phase behavior.

The coefficients of the expansion depend on the temperature
and can be expanded themselves about the critical temperature.
Using Eqs.\  (\ref{MSpinodal}) and (\ref{MCritPoint}), and
keeping the lowest order term for each coefficient, we obtain:

\begin{eqnarray}
\label{MEnergy2}
F(\rho,T) \approx 
\frac{1}{2}\,\left.\frac{\partial^ 3F}{\partial\rho^2\,\partial T}
\right|_{c}\delta T\,{\delta\rho}^2 +\phantom{\hspace{1cm}} \nonumber \\
\frac{1}{3!}\,\left.\frac{\partial^ 4F}{\partial\rho^3\,\partial T}
\right|_{c}\delta T\,{\delta\rho}^3 + 
\frac{1}{4!}\,\left.\frac{\partial^ 4F}{\partial\rho^4}
\right|_{c}{\delta\rho}^4 + \dots,
\end{eqnarray}
where $\delta T\equiv\,T-T_c$. Since $\delta T \sim \delta\rho^2$ (see Eq.\ (\ref{MCoex}) below) the term in $\delta T{\delta\rho}^3$ 
is of higher order and can be neglected. Therefore, the lowest 
order expansion of the free energy has the simple form:

\begin{eqnarray}
\label{MEnergy3}
F(\rho,T) \approx \;
A\;\delta T\,{\delta\rho}^2 + C\;\,{\delta\rho}^4,
\end{eqnarray}
for appropriate $A$ and $C$. The coexisting phases
obey the standard result
\begin{eqnarray}
\label{MCoex}
\delta T\approx -\,\frac{2\,C}{A}\;\delta\rho^2,
\end{eqnarray}
giving the well-known parabolic shape of the phase
diagram around the critical point for monodisperse
systems (always within the mean field approximation).

\section{Polydisperse System}

In the polydisperse case, the particles are not all identical but 
differ in certain scalar attributes 
(particle radius, charge, chain length...) which can be tagged with indices 
$\sigma_1,\sigma_2,\dots$. For simplicity, we
focus on systems with just one polydisperse index $\sigma$,
and write the corresponding density of particles as
$\rho(\sigma)$. In this case, the free energy 
can be written as two parts, an ideal term and an excess
term, as follows \cite{Ideal}:
\begin{eqnarray}
\label{Energy00}
F[\rho]=\int\!\!d{\sigma}\,\;
\rho(\sigma)\left\{\log\left(\rho(\sigma)\right)-1\right\} + 
F_{ex}[\rho].
\end{eqnarray}

While the ideal part only depends on $\rho(\sigma)$ as shown, the excess part 
$F_{ex}$ is, in general, an arbitrary functional of $\rho(\sigma)$ and a 
function of $T$. Here, in order to explore how polydispersity affects 
the Landau expansion around the critical point in a tractable manner, 
we restrict ourselves to cases where the functional dependence of the
excess part of the free energy on the density
distribution $\rho(\sigma)$ is through a single 
`generalized moment' \cite{Projected} of the distribution:

\begin{eqnarray}
\label{Energy0}
F[\rho]=\int\!\!d{\sigma}\,\;
\rho(\sigma)\left\{\log\left(\rho(\sigma)\right)-1\right\} + f(\rho_1,T),
\end{eqnarray}
where 
\begin{eqnarray}
\label{Moment1}
\rho_1 = \int\!\!d{\sigma}\,\;
\rho(\sigma)\, w_1(\sigma)
\end{eqnarray}
with $w_1(\sigma)$ an arbitrary weight function.
Although this assumption seems restrictive, it is true for some important 
models including the Flory-Huggins theory of length-polydisperse polymer 
solutions and blends \cite{Projected}.
The models for which the excess part of the free
energy depends on the density $\rho(\sigma)$ through
a finite set of moments of this distribution, for
which Eq.\ (\ref{Energy0}) is a particular example,
are usually termed {\it truncatable} models.

\subsection{The Parent Distribution Function}

Polydispersity typically appears in the process of chemical
manufacture of the particles of the system, so that the relative proportion of particles with different $\sigma$ is fixed thereafter. The actual density of each species then depends 
on how much solvent is present. We define a parent distribution $P(\sigma)$ as
\begin{eqnarray}
\label{Dilution}
P(\sigma)=\,P_0\;n(\sigma),
\end{eqnarray}
where $n(\sigma)$ is a normalized distribution (fixed by the initial preparation 
of the system) and $P_0$ is the overall density of particles, which varies as one moves along a `dilution line' through the phase diagram.
All other moments
of the parent are fixed by $P_0$; for example, 
\begin{eqnarray}
\label{PMoment1}
P_1 = \int\!\!d{\sigma}\,\;
P(\sigma)\, w_1(\sigma) = \,
P_0\;n_1
\end{eqnarray}
where $n_1$ is fixed at synthesis via the parent: \begin{eqnarray}
\label{PMoment2}
n_1 = \int\!\!d{\sigma}\,\;n(\sigma)\, w_1(\sigma).
\end{eqnarray}

Although the parent distribution is globally fixed, the system can separate into coexisting phases with a distribution $\rho(\sigma)$ that is locally different from $P(\sigma)$. If there are two such phases
$\alpha$ and $\beta$, with respective distributions $\rho^{\alpha}(\sigma)$ 
and $\rho^{\beta}(\sigma)$, the parent distribution is recovered as the volumetric average:
\begin{eqnarray}
\label{tieline}
x^{\alpha}\,\rho^{\alpha}(\sigma)\,+\,x^{\beta}\,\rho^{\beta}(\sigma)
\,=\,P(\sigma),
\end{eqnarray}
where $Vx^{\alpha}$ and $Vx^{\beta}$ are the volumes of
phases $\alpha$ and $\beta$, respectively. 
Note however that at a phase boundary which marks the onset of phase separation, the volume fraction of the incipient (shadow) phase is 
negligible and, therefore, the $\sigma$-distribution of the original (cloud) 
phase must coincide with the parent distribution.

\subsection{Projection Method}

Although for a given $n(\sigma)$ we require only one density (such as $\rho_1$),
 alongside temperature, to specify a point on the phase diagram, the 
corresponding free energy must depend on the full density
distribution $\rho(\sigma) = P_0 n(\sigma)$. Accordingly it is not correct to 
substitute the parent distribution in a functional like Eq.\ (\ref{Energy0}) to
 obtain a one-variable free energy $F(P_0)$ (or equivalently, $F(\rho_1)$). 
In fact such a substitution would describe a case where the normalized 
distribution $n(\sigma)$ were constrained to be equal in all coexisting 
phases -- which is not the physical situation. Instead,
in the case of Eq.\ (\ref{Energy0}),
the equality of chemical potentials dictates that
the $\sigma$-distribution of any cloud phase (which coexists, by definition, 
with an unperturbed parent) must obey
\begin{eqnarray}
\label{CoexistDF}
\rho(\sigma)\,=\,P(\sigma)\;e^{\;\lambda\,w_1(\sigma)},
\end{eqnarray}
for an appropriate value of $\lambda$. This distribution
is not proportional to the parent unless $\lambda = 0$ (when the phases are 
identical).

However, under certain conditions there is a 
procedure to transform the free energy functional into an effective free energy 
{\it function}. The conditions are first that (as in Eq.\ (\ref{Energy0})) the 
excess free 
energy $f$ only depends on the distribution $\rho(\sigma)$ through a finite set 
of moments (defined as in Eq.\ (\ref{Moment1})); and second that
we are only interested in phase 
boundaries (the cloud and the shadow curves) and the
spinodals. The procedure, called the projection
method \cite{Projected,Review}, involves several steps briefly described here, 
and leads to a free energy function dependent only on the moments required to 
specify the excess free energy.

First, considers the free energy

\begin{eqnarray}
\label{Energy}
F[\rho]=\int\!\!d{\sigma}\,\;
\rho(\sigma)\left\{\log\left(\frac{\rho(\sigma)}
{P(\sigma)}\right)-1\right\} + f(\rho_1,T),
\end{eqnarray}
which, for our purposes, is equivalent to Eq.\ (\ref{Energy0}),
since the inclusion of the parent as a denominator in the logarithm represents a linear term (which can be considered as a mere
shift of the chemical potentials). But note that, for that equivalence to hold,
the parent $P(\sigma)$, including $P_0$, must be kept fixed 
as the density distribution $\rho(\sigma)$ varies: this observation is crucial to the Landau expansion developed below.

The projection method follows
by minimising the first (ideal) term of this free energy subject to the constraint of fixed values for the moment(s) upon which the excess free energy depends. (The only such moment is $\rho_1$ in our case.) This yields 

\begin{eqnarray}
\label{ProjDist}
\rho(\sigma)= P(\sigma)\;
e^{\,\lambda_1\,w_1(\sigma)},
\end{eqnarray}
where $\lambda_1$ is the Lagrange multiplier
used to constrain the minimization, determined by $\rho_1$ through
\begin{eqnarray}
\label{Rho1}
\rho_1=\int\!\!d{\sigma}\,\; P(\sigma)\,w_1(\sigma)\,
e^{\,\lambda_1\,w_1(\sigma)}.
\end{eqnarray}
Since both the parent phase (the cloud) and the coexisting phase 
(the shadow, Eq.\ (\ref{CoexistDF})) obey Eq.\ (\ref{ProjDist})
for different values of $\lambda_1$, the method contains no approximation
when it comes to finding the phase boundaries. However, if the $\sigma$-distribution of 
any of the phases adopts a form different from (\ref{ProjDist}),
the method produces only approximate results.
This occurs whenever the system separates into finite volumes of more than one phase, that is, everywhere {\it within} the cloud curve. (In Eq.\ (\ref{tieline}),
the projection method is inexact when $x_\alpha$ and $x_\beta$
are both nonzero.)

The free energy obtained by this method is \cite{Projected}
\begin{eqnarray}
\label{Energy2}
F(\rho_1,T)=\lambda_1\,\rho_1 - \rho_0 + f(\rho_1,T),
\end{eqnarray}
where 
\begin{eqnarray}
\label{Rho0}
\rho_0=\int\!\!d{\sigma}\,\; P(\sigma)\,
e^{\,\lambda_1\,w_1(\sigma)}.
\end{eqnarray}
This free energy depends on only
one density variable, since both $\lambda_1$ and $\rho_0$
are given by $\rho_1$. The coexisting phases are obtained from 
this free energy {\it function} via the common tangent method,
as in the monodisperse case.

\subsection{Dilution Line}

As mentioned above, the projected free energy requires
knowledge of the parent distribution function 
$P(\sigma)$, whose normalization 
$P_0$ varies as one moves along a dilution line on the phase diagram.
For this reason, the projected free energy $F(\rho_1,T)$ depends
parametrically on $P_0$,
through $\lambda_1(\rho_1,P_0)$ 
and $\rho_0(\lambda_1,P_0)$; see Eqs.\
(\ref{Rho1}) and (\ref{Rho0}), respectively.
This dependence is parametric because, in the process of
calculating the coexistence between phases, $P_0$ must remain fixed, so as to maintain the equivalence of Eq.\ (\ref{Energy0}) and
(\ref{Energy}).

Suppose that, for fixed $P_0$ and $T$, we obtain by this method a 
coexistence between phases $\alpha$ and $\beta$, with moment densities $\rho^{\alpha}_1$
and $\rho^{\beta}_1$, respectively. This determines the value 
of $\lambda^{\alpha}_1$, $\lambda^{\beta}_1$, $\rho^{\alpha}_0$, and 
$\rho^{\beta}_0$ through Eqs.\ (\ref{Rho1}) and (\ref{Rho0})
which, in turn, determines the density distribution of the two
phases $\rho^{\alpha}(\sigma)$ and $\rho^{\beta}(\sigma)$ via
Eq.\ (\ref{ProjDist}). In general, neither of these distributions
is the parent distribution. 

This is consistent with the fact that 
the parent distribution does not coexist with a second phase for arbitrary combinations 
of $P_0$ and $T$; this happens only on the cloud curve.
To find the cloud curve, we combine the coexistence conditions derived above with the equation of the dilution line, so that one of the coexisting phases ($\alpha$ say)
is the parent distribution; this implies that 
$\rho^{\alpha}(\sigma)\!=\!P(\sigma)$, 
$\lambda^{\alpha}_1\!=\!0$, $\rho^{\alpha}_0\!=\!P_0$, and
$\rho^{\alpha}_1\!=\!P_0\,n_1$ (or $\rho^{\alpha}_1\!=\!\rho^{\alpha}_0\,n_1$).

\subsection{Critical Point}

The critical point of the polydisperse system lies at the
intersection of the cloud and shadow curves and is, therefore,
correctly described (given Eq.\ (\ref{Energy0})) by the projected free energy. The
equations to determine it are analogous to those of the
monodisperse case,
\begin{eqnarray}
\label{Spinodal}
\left.\frac{\partial^ 2F}{\partial\rho_1^2}\right|_{c} = 0
\end{eqnarray}
and
\begin{eqnarray}
\label{CritPoint}
\left.\frac{\partial^
3F}{\partial\rho_1^3}\right|_{c} = 0,
\end{eqnarray}
where $|_c$ means evaluated at the critical point $(\rho_1^c,T_c;P_0^{\,c})$.
Note that these two equations have three unknowns, namely
$\rho_1^c$, $T_c$ and $P_0^{\,c}$, and must be complemented
by the dilution line constraint, $\rho_1^c\!=\!P_0^{\,c}\,n_1$.

\section{Landau Expansion}

The projected free energy $F(\rho_1,T;P_0)$
can now be expanded about the critical point,

\begin{eqnarray}
\label{Energy3}
F\approx 
\frac{1}{2}\,\left.\frac{\partial^ 2F}{\partial\rho_1^2}
\right|_{\rho_1^c,T;P_0}{\delta\rho}^2 + \phantom{\hspace{1cm}} \nonumber \\
\frac{1}{3!}\,\left.\frac{\partial^ 3F}{\partial\rho_1^3}
\right|_{\rho_1^c,T;P_0}{\delta\rho}^3 + 
\frac{1}{4!}\,\left.\frac{\partial^ 4F}{\partial\rho_1^4}
\right|_{\rho_1^c,T;P_0}{\delta\rho}^4 + \dots,
\end{eqnarray}
where $\delta\rho\equiv\,\rho_1-\rho_1^c$. As before, the 
constant and linear terms have been dismissed.

Now we must expand the coefficients of this
expansion around the critical values not only of $T$, as we did in the monodisperse case, but also of 
$P_0$.
One might think that, after doing this, the free energy could
be written, to the lowest order, as
\begin{eqnarray}
\label{Energy5}
F\approx \left(A_1\,\delta T + A_2\,\delta P_0
\right)\,{\delta\rho}^2 + C\,{\delta\rho}^4,
\end{eqnarray}
where $\delta P_0\equiv\,P_0-P_0^{\,c}$. If so,
the term in brackets should be of order ${\delta\rho}^2$
to balance the term of order ${\delta\rho}^4$. However,
the second term in the brackets is actually linear in ${\delta\rho}$,
as dictated by the dilution line $\delta P_0 = \delta\rho/n_1$.
These two facts can only be obeyed if the first term in the brackets 
behaves like
$A_1\delta T\approx-A_2\,\delta\rho/n_1+{\cal O}\left({\delta\rho}^2\right)$ 
(instead of  $\delta T\sim{\delta\rho}^2$, as in the monodisperse case).
This corresponds to the fact that, generically, the critical point 
is no longer at the top of the coexistence curve. Therefore,
to maintain consistently the order of the approximation, the correct 
lowest order expansion of the free energy must contain all terms
of order ${\delta\rho}^4$:

\begin{eqnarray}
\label{Energy4}
F\approx 
\frac{1}{2}\left[
\left.\frac{\partial^ 3F}{\partial\rho_1^2\partial T}\right|_{c}
\!\delta T
+
\left.\frac{\partial^ 3F}{\partial\rho_1^2\partial P_0}\right|_{c}
\!\delta P_0
+ 
\frac{1}{2}\left.\frac{\partial^ 4F}{\partial\rho_1^2\partial T^2}\right|_{c}
\!{\delta T}^2
+\right. \nonumber \\
\left.
\frac{1}{2}\left.\frac{\partial^ 4F}{\partial\rho_1^2\partial P_0^{\,2}}\right|_{c}
\!{\delta P_0}^2
+ 
\left.\frac{\partial^ 4F}{\partial\rho_1^2\,\partial T\,\partial P_0}\right|_{c}
\!\delta T\;\delta P_0
\;\;\right]{\delta\rho}^2 
+ \nonumber \\
\frac{1}{3!}\left[
\left.\frac{\partial^ 4F}{\partial\rho_1^3\partial T} \right|_{c} 
\!\delta T
+
\left.\frac{\partial^ 4F}{\partial\rho_1^3\partial P_0} \right|_{c} 
\!\delta P_0
\right]{\delta\rho}^3
+ \nonumber \\
\frac{1}{4!}\left.\frac{\partial^ 4F}{\partial\rho_1^4} \right|_{c}
{\delta\rho}^4 + \dots .
\end{eqnarray}

To continue, let us define the functions
\begin{eqnarray}
\label{Moments}
\rho_m = \int\!\!d{\sigma}\,\;
P(\sigma)\, \left(w_1(\sigma)\right)^m
\;e^{\,\lambda_1\,w_1(\sigma)},
\end{eqnarray}
for any $m=0,1,2,\dots$, which are an expanded set of generalised
moments consistent with the previous definitions 
Eq.\ (\ref{Rho1}) and (\ref{Rho0}). These functions depend
on $\rho_1$ and $P_0$ and, indeed, they obey:

\begin{eqnarray}
\label{PropRho1}
\frac{\partial\rho_m}{\partial\rho_1} =
\frac{\rho_{m+1}}{\rho_2},
\end{eqnarray}
and
\begin{eqnarray}
\label{PropRho2}
\frac{\partial\rho_m}{\partial P_0} =
\frac{1}{P_0}
\left(\rho_m-\frac{\rho_1\rho_{m+1}}{\rho_2}\right).
\end{eqnarray}

Very importantly, these equations imply that the derivatives of any
(generalised) moment with respect to $\rho_1$ gives rise to functions
of moments (Eq.\ (\ref{PropRho1})), and the derivatives of any moment with 
respect to $P_0$ must vanish in the monodisperse limit 
(as does Eq.\ (\ref{PropRho2})).

With these definitions, we can write the second derivative of 
the free energy, from which all the coefficients of the expansion 
(\ref{Energy4}) can be obtained, as:

\begin{eqnarray}
\label{F2}
\frac{\partial^2 F}{\partial{\rho_1}^2} =
\frac{1}{\rho_2}+\frac{\partial^2 f(\rho_1,T)}{\partial{\rho_1}^2}.
\end{eqnarray}
Now note that, since $T$ and $P_0$ appear in disjunct 
parts of the free energy (as in the original free energy, 
Eq.\ (\ref{Energy2})), all crossed terms must vanish:

\begin{eqnarray}
\label{Prop0}
\frac{\partial^ 2F}{\partial T\,\partial P_0} =0.
\end{eqnarray}
Accordingly, the coefficients of the expansion can be divided into
three different groups: those including only derivatives with respect to
$\rho_1$, those including derivatives with respect to
$T$, and those including derivatives with respect to $P_0$. 
While the first group of coefficients originates from the
full free energy, the second originates from the excess part of 
the free energy (`the interactions'), and the third
group arises from the ideal part of the free energy (`the entropy').
These three groups of coefficients have a distinct dependence on polydispersity.
In particular, all the coefficients stemming from the ideal part 
(those multiplying any power 
of $\delta P_0$) must vanish in the monodisperse limit \cite{Vanish}, 
hence recovering the ordinary Landau expansion, Eq.\ (\ref{MEnergy3}) 
(plus some terms e.g. in $\delta T^2 \delta \rho^2$ that are higher order 
corrections in this limit). 

In brief, the correct lowest order expansion of the free energy, 
retaining  all terms to fourth order in density, is therefore
\begin{eqnarray}
\label{Energy6}
F\approx \left(A_1\,\delta T + A_2\,\delta P_0 +
A_3\,{\delta T}^2 + A_4\,{\delta P_0}^2\right)\;{\delta\rho}^2 +
\nonumber \\
\left(B_1\,\delta T + B_2\,\delta P_0\right)\;{\delta\rho}^3
+ C_0\;{\delta\rho}^4 \phantom{\hspace{1cm}},
\end{eqnarray}
where the coefficients of terms involving $\delta T$ (those 
with odd indices: $A_1$, $A_3$ and $B_1$) only depend on
derivatives of the excess free energy $f$, and the coefficients 
of terms involving $\delta P_0$ ($A_2$, $A_4$ and $B_2$) stem solely from 
the ideal entropy term of the system and not the interactions. The latter
vanish in the monodisperse limit; for example,
\begin{eqnarray}
\label{CoefA2}
A_2\equiv\;\frac{1}{2\,{P_0^c}^{\,2}}\,\left(
\frac{n_1\,n_3}{{n_2}^3}-
\frac{1}{n_2}\right),
\end{eqnarray}
where $n_1,\,n_2\dots$ are generalized moments of the 
(normalized) parent distribution,
\begin{eqnarray}
n_m\equiv \int\!\!d{\sigma}\,\;
n(\sigma)\, \left(w_1(\sigma)\right)^m,
\end{eqnarray}
which do not depend on any thermodynamic variable.

It is interesting to point out here that the ultimate
reason why the critical point is not on the top of
the cloud curve is that the squared term of the free energy 
expansion at the critical point behaves effectively as
${\delta T}^{\,2}\,{\delta\rho}^{\,2} $ instead of the usual
${\delta T}\,{\delta\rho}^{\,2}$.

We stress that although the coefficients $A_2$, $A_4$ and $B_2$ 
are small for narrow polydispersities, the expansion 
(\ref{Energy6}) is asymptotically exact around the critical point for arbitrary 
(not necessarily narrow) polydisperse distributions.
As ever, `exact' means exact within a mean-field free energy of the form Eq.\ (\ref{Energy0}).

\subsection{Cloud Curve}

\label{cloudsec}
In order to obtain an expression for the cloud curve,
we can write the free energy expansion (\ref{Energy6}) in terms 
of a {\it shifted} density

\begin{eqnarray}
\label{ShifVar}
\delta\eta\equiv\delta\rho+\frac{B_1\,\delta T +
B_2\,\delta P_0}{4\,C_0},
\end{eqnarray}
obtaining an asymptotic expression of the form

\begin{eqnarray}
\label{Energy7}
F\approx A\;{\delta\eta}^2 + C_0\;{\delta\eta}^4,
\end{eqnarray}
where
\begin{eqnarray}
\label{defA}
A\equiv A_1\,\delta T + A_2\,\delta P_0 +
A_3\,{\delta T}^2 + A_4\,{\delta P_0}^2 
\nonumber \\
- \frac{3}{8}\frac{(B_1\,\delta T + B_2\,\delta P_0)^2}{C_0}.
\phantom{\hspace{1.5cm}}
\end{eqnarray}

Starting from this simple expression, the coexisting states follow
immediately as
\begin{eqnarray}
\label{Coex1}
\delta\eta\approx \pm\sqrt{-\frac{A}{2\,C_0}}.
\end{eqnarray}
These two coexisting densities correspond to distribution functions
given by Eq.\ (\ref{ProjDist}), neither of which is in general
the parent distribution function $P(\sigma)$. 
In order to obtain the appropriate cloud curve, we must therefore impose 
the dilution line constraint, 
$\delta\rho^{\scriptscriptstyle CL}=n_1\,\delta P_0$,
which leads to the following expression for the cloud curve:

\begin{eqnarray}
\label{Cloud}
\delta T_{\scriptscriptstyle CL} \approx \alpha_{\scriptscriptstyle CL}\;
{\delta\rho} + \beta_{\scriptscriptstyle CL}\;{\delta\rho}^2
+ \;\,{\cal O}({\delta\rho}^3),
\end{eqnarray}
where the slope of the cloud curve at the critical point is given by
\begin{eqnarray}
\label{alphacl}
\alpha_{\scriptscriptstyle CL} \equiv -\,\frac{A_2}{A_1\,n_1},
\end{eqnarray}
and its curvature by
\begin{eqnarray}
\label{betacl} 
\beta_{\scriptscriptstyle CL} \equiv 
\frac{C_0}{A_1}\left[\,\left(1-\Delta\right)^{2}\!-3\,\right]
-\frac{A_4}{A_1\,n_1^2}-\frac{A_2^2\,A_3}{A_1^3\,n_1^2},
\end{eqnarray}
where 
\begin{eqnarray}
\label{Delta}
\Delta\equiv\frac{\,A_1B_2\!-\!A_2B_1}{2\,A_1\,C_0\,n_1\,}.
\end{eqnarray}
As predicted (preceding Eq.\ (\ref{Energy4}) above), we have thus established that
$A_1\,\delta T + A_2\,\delta P_0 \sim {\delta\rho}^2$,
with exact cancelation of the terms linear in $\delta\rho$. This
shows the consistency of the approximation and proves that none of the
terms in Eq.\ (\ref{Energy4}) can be discarded in the expansion.

One consequence of these results concerns the geometric location of the critical point 
along the cloud curve. For the critical point to remain at the apex of the 
cloud curve (which is where it resides in the monodisperse limit),
the slope of the cloud curve at the critical point 
$\alpha_{\scriptscriptstyle CL}$, Eq.\ (\ref{alphacl}), must vanish;
this requires $A_2=0$. According to Eq.\ (\ref{CoefA2}), this requires very 
particular combinations of the parent distribution and weight function 
$w_1(\sigma)$. If the latter only adopts positive values, 
$A_2$ is strictly positive except in the monodisperse limit so that the 
critical point will move away from the top of the cloud curve even for 
infinitesimal polydispersity \cite{sign}. Since the argument originates 
from the entropic part of the free energy, it may hold more generally than 
for systems obeying Eq.\ (\ref{Energy0}) \cite{caveat1,Baus}.

For monodisperse systems, the slope $\alpha_{\scriptscriptstyle CL}$ 
vanishes, while $\beta_{\scriptscriptstyle CL}$ tends to $-{2C_0}/{A_1}$, 
thus recovering the classical scenario that leads to Eq.\ (\ref{MCoex}): 
the cloud curve forms the binodal curve with a quadratic maximum where the 
critical point resides.

\subsection{Shadow Curve}

The shadow curve does not, of course, obey the dilution line constraint,
and therefore is not obtainable direct from Eq.\ (\ref{Coex1}).  Nonetheless we expect \begin{eqnarray}
\label{Shadow}
\delta T_{\scriptscriptstyle SH} \approx \alpha_{\scriptscriptstyle SH}\;{\delta\rho} +
 \beta_{\scriptscriptstyle SH}\;{\delta\rho}^2
+ \;\,{\cal O}({\delta\rho}^3).
\end{eqnarray}

To find $\alpha_{\scriptscriptstyle SH}$, we may use Eq.\ (\ref{Coex1}) to 
write a relation between the cloud and the shadow curves,
\begin{eqnarray}
\label{Coex2}
\delta\eta^{\scriptscriptstyle CL} + \delta\eta^{\scriptscriptstyle SH} 
\approx 0.
\end{eqnarray}
This yields, upon substitution of Eq.\ (\ref{ShifVar}), 
\begin{eqnarray}
\label{alphash}
\alpha_{\scriptscriptstyle SH} \equiv 
-\,\frac{\alpha_{\scriptscriptstyle CL}}{\;1\,+\,\Delta\;}.
\end{eqnarray}

Interestingly though, Eq.\ (\ref{Coex2}) does not provide the correct curvature of the shadow curve, $\beta_{\scriptscriptstyle SH}$, as can easily be checked numerically for the example considered below (Section \ref{examplesec}).
This is because the curvature of the shadow in fact depends on some 
higher-order terms (of order $\delta\rho^5$) that were not included in 
Eq.\ (\ref{Energy6}). The fifth-order terms are
\begin{eqnarray}
\label{Energy8}
\nonumber 
\Delta F = \left(A_5\,{\delta T}^3 + A_6\,{\delta P_0}^3\right)\;
{\delta\rho}^2 +\left(B_3\,{\delta T}^2 + B_4\,{\delta P_0}^2\right)\;
{\delta\rho}^3 \hspace*{-1.5cm}\\ 
+\left(C_1\,\delta T + C_2\,\delta P_0\right)\;{\delta\rho}^4
+ D_0\;{\delta\rho}^5,
\end{eqnarray}
Including these (fifth-order)
terms in the expansion, Eq.\ (\ref{Energy6}),
transforms Eq.\ (\ref{Coex2}) into
\begin{eqnarray}
\label{Coex3}
\delta\eta^{\scriptscriptstyle CL} + \delta\eta^{\scriptscriptstyle SH} 
\approx \phantom{\hspace{2.5cm}}
\\ \nonumber
\frac{(A_1\,\delta T + A_2\,\delta P_0 +
A_3\,{\delta T}^2 + A_4\,{\delta P_0}^2)\,D_0}{2\,\left(
C_0+C_1\,\delta T + C_2\,\delta P_0\right)^2}+{\cal O}({\delta\eta}^3),
\end{eqnarray}
from which we finally obtain the expresion
\begin{eqnarray}
\label{betash} 
\beta_{\scriptscriptstyle SH} \equiv 
\frac{1}{\left(\;1\,+\,\Delta\;\right)^{3}}\,\left[
\beta_{\scriptscriptstyle CL}
\left(1+\frac{1}{2\,C_0^2\,n_1}(B_2 C_0-A_2 D_0)\right) 
\right.\nonumber \\ 
+\frac{\alpha_{\scriptscriptstyle CL}}{2\,C_0^2\,n_1^2}
(B_2 C_2-B_4 C_0 + A_4 D_0)
\nonumber \\ 
+\frac{\alpha_{\scriptscriptstyle CL}^{\,2}}{2\,C_0^2\,n_1}
(B_1 C_2+B_2 C_1)
\nonumber \\ \left.
+\frac{\alpha_{\scriptscriptstyle CL}^{\,3}}{2\,C_0^2}
(B_1 C_1-B_3 C_0+A_3 D_0)
\right].
\end{eqnarray}
In the monodisperse limit, analogously to the cloud curve, the slope 
$\alpha_{\scriptscriptstyle SH}$ vanishes and  $\beta_{\scriptscriptstyle SH}$ 
tends to $-{2C_0}/{A_1}$, thus forming the binodal curve by the merger of cloud and shadow.

\subsection{Consequences of the Expansion}

Note the progression in complexity of the coefficients needed in the
expansion to compute the slopes, and then curvatures, of the cloud and shadow at the critical point. The slope of the cloud curve $\alpha_{\scriptscriptstyle CL}$ 
depends on just two coefficients of the expansion ($A_1$ and $A_2$);
the slope of the shadow involves three extra coefficients 
($B_1$, $B_2$ and $C_0$); the curvature of the cloud at the 
critical point requires knowledge of all seven
coefficients of the quartic Landau expansion, Eq.\ (\ref{Energy6}); 
that of the shadow also involves a selection of the fifth order terms 
(but not all of them): Table 1 shows which terms in the Landau expansion 
must be retained to correctly predict each of these parameters.

Although these results are obtained for a one-moment free-energy, 
Eq.\ (\ref{Energy0}), they involve higher moments of the parent distribution 
\cite{Projected}, as detailed in Table 1. Note also that the critical point and 
the spinodals depend only on a reduced set of moments; this holds to all
orders in the Landau expansion. In contrast, order by order, the cloud and 
shadow curves involve progressively all the moments of the parent distribution, 
through Eqs.\ (\ref{PropRho1}) and (\ref{PropRho2}).

The results given above were obtained by the projection method, but since this 
gives the exact cloud and shadow curves, any other method for finding
$\alpha_{\scriptscriptstyle CL}$, $\beta_{\scriptscriptstyle CL}\dots$ 
within mean field theory must produce, necessarily, expressions equivalent to 
Eqs.\ (\ref{alphacl},\ref{betacl},\ref{alphash},\ref{betash}).

\subsection{Phase Diagram in Temperature-Pressure Plane}

In the monodisperse case, the phase equilibria show up
in the $T\!-\!p$ phase diagram as a line of coexistence, 
ending at the critical point. If polydispersity is
included, this line now broadens and adopts a banana-like shape 
of finite area, containing the values of temperature and pressure
for which the system undergoes phase separation. In this case, the critical 
point appears on the perimeter of the coexistence region. 

We now can calculate the shape of the coexistence region to the same order
of approximation as we did with the cloud and the shadow. After projection of 
the free energy to give Eq.\ (\ref{Energy2}), the pressure of the system can be 
written in the familiar form \cite{Projected}
\begin{eqnarray}
\label{Press}
\frac{p}{k_{B}T}=\frac{\partial F}{\partial\rho_1}\;\rho_1\,-\,F,
\end{eqnarray}
which allows us to expand $\delta p\equiv p-p_c$ as
\begin{eqnarray}
\label{Press2}
\frac{\delta\,p}{k_{B}T}\approx
a_1\,\delta T + a_2\,\delta P_0 +
b_1\,{\delta T}^2 + b_2\,{\delta P_0}^2+\dots
\end{eqnarray} 
where (just as occurred in Eq.\ (\ref{Energy6})) the coefficients of terms 
involving $\delta T$ (those 
with odd indices: $a_1$ and $b_1$) only depend on
derivatives of the excess free energy $f$, and the coefficients 
of terms involving $\delta P_0$ ($a_2$ and $b_2$) derive solely from 
the ideal entropy of the system (not the interactions) and vanish in the 
monodisperse limit. For instance,
\begin{eqnarray}
\label{a2}
a_2\equiv\;\frac{n_1^{\,2}}{n_2}-1\;\;(\,\le 0\,).
\end{eqnarray}

Substitution of the approximation derived in Section \ref{cloudsec} for the cloud curve
yields
\begin{eqnarray}
\label{Press3}
\frac{\delta\,p}{k_{B}T}\approx
\alpha_{\scriptscriptstyle P}\;{\delta\rho} +
 \beta_{\scriptscriptstyle P}\;{\delta\rho}^2,
\end{eqnarray} 
where
\begin{eqnarray}
\label{alphap}
\alpha_{\scriptscriptstyle P} \equiv 
a_1\,\alpha_{\scriptscriptstyle CL}\,+\,\frac{a_2}{n_1}
\end{eqnarray}
and
\begin{eqnarray}
\label{betap} 
\beta_{\scriptscriptstyle P} \equiv 
a_1\,\beta_{\scriptscriptstyle CL}\,+\,
b_1\,{\alpha_{\scriptscriptstyle CL}}^2\,+\,
\frac{b_2}{n_1^2}.
\end{eqnarray}

A generic consequence of these results concerns the location of the critical point along
the perimeter of the coexistence region on the $T\!-\!P$ phase diagram.
It is clear on geometrical grounds that the critical point 
cannot be located at the point of maximum temperature and maximum pressure
simultaneously, unless the coexistence region shows a kink at that point.
Such a kink is not predicted by the Landau expansion, which gives smooth 
behavior everywhere; nor have we seen evidence for it in any of the models 
(or experiments) studied to date. Assuming these two turning points are indeed distinct,
 one can ask whether the critical point generically lies at one, at the other, 
or neither. We argued already above that generically, for systems obeying 
Eq.\ (\ref{Energy0}), the critical point is not at the maximum of $T$ since 
this would require $A_2$ to vanish (Eq.\ \ref{CoefA2}). Likewise, for it to 
lie at the pressure maximum
would require $\alpha_{\scriptscriptstyle P}=0$ in Eq.\ (\ref{alphap}), 
for which a fine tuning of the parameters appears necessary \cite{caveat2,Baus}.
We conclude that generically it lies at neither of these points; indeed, 
as shown in the following example, it lies between them.

\section{Example}
\label{examplesec}
In this Section, we illustrate the results of the Landau
expansion for the cloud and shadow curves of a particular
system: an incompressible blend of two homopolymers.
One of the polymers is considered polydisperse
in length, with number densities $\rho(\sigma)$ where $\sigma$ is the number 
of monomers on a chain (effectively a continuous parameter). The other polymer 
(or solvent) is monodisperse and
has $N'$ monomers. For this system, the free energy can be
easily obtained from the Flory-Huggins theory
and recovers the form (\ref{Energy0}). Specifically, the excess free energy
reads \cite{Projected}
\begin{eqnarray}
\label{FHf}
f(\rho_1,\chi)=\frac{1-\rho_1}{N'}\log\left(\frac{1-\rho_1}{N'}\right)
+ \chi\,\rho_1\,(1-\rho_1),
\end{eqnarray}
where $\chi$ is the Flory parameter, which plays the role of an 
inverse temperature. The weight function for the moment $\rho_1$
is $w_1(\sigma)=\sigma$ (in the usual polymer jargon, $\rho_1$
is the monomer volume fraction $\phi$).

As a normalised parent, we consider in this example a uniform
distribution
\begin{equation}
\label{h}
n(\sigma)= \left\{ \begin{array}{cl}
\frac{\textstyle 1}{\;\textstyle 2\sqrt{3}\,\Delta N\,} & \textrm{  if $|\sigma-N|\le \sqrt{3}\,\Delta N$}\\
 & \\
0 & \textrm{  otherwise}
\end{array}\right.
\end{equation}
where $N=\!n_1\!=\langle\sigma\rangle$ is the mean polymer size,
and $\Delta N\!=\!(\langle\sigma^2\rangle-\langle\sigma\rangle^2)^{1/2}$
is the standard deviation.

We plot in Figures 1 and 2 the resulting behavior for the 
case where the polymer solvent has $N'\!=\!10$ monomers and the 
distribution of sizes of the solute follows Eq.\ (\ref{h}) with $N=1000$
and $\Delta N/N\!=\!0.2$. (This corresponds, in polymer jargon, 
to a polydispersity index 
$M_w/M_n\!=\langle\sigma^2\rangle/\langle\sigma\rangle^2 = 1.4$.)

We see from these plots that the Landau expansion, Eq.\ (\ref{Energy6}), 
captures correctly all the essential features of the full mean-field model near 
the critical region, just as it does for monodisperse systems. 
In this example, the Landau predictions cross the cloud and shadow curves at the critical point (although the slopes and curvatures are correct). This behavior is inherited from the monodisperse case, and is presumably connected with an error in the third derivative of these curves.

Given the increased complexity of the phase behavior (separation of cloud and shadow 
curves; critical point not at the maximum of these; finite width coexistence 
region on the $p-T$ plot collapsing to a line in the monodisperse limit), we 
find this somewhat remarkable. We emphasize that every term in 
Eq.\ (\ref{Energy6}) must be retained to achieve this degree of consistency -- 
and in particular to reproduce the correct curvature of the cloud;
to get the curvature of the shadow also correct, one needs as well the higher terms in Eq. (\ref{Energy8}). The latter were included in the figure although leaving them out gives a result that, for this example, is only slightly less accurate. Despite all the extra complexity, the region 
of the phase diagram within which the Landau expansion could tolerably be 
used to replace the full calculation is similar to the monodisperse case.

\section{Conclusions}

Within the one-moment form of the excess free energy represented by 
Eq.\ (\ref{Energy0}), we have derived a generalized Landau 
expansion, Eqs.\ (\ref{Energy6},\ref{Energy8}), from which the characteristic features of 
polydisperse binary fluid coexistence can all be shown to follow. This 
Landau expansion was compared with direct calculation from 
the underlying (mean-field) free energy in a specific example (Figures 1 
and 2) with excellent agreement near the critical point and qualitative 
agreement away from it. Note that truncation of the Landau expansion at the quartic order (Eq.\ref{Energy6}) may prove adequate for qualitative purposes if the curvature of the shadow is not important.  

Our expansion is correct around the critical point for arbitrary 
polydispersity, be it broad or narrow. The monodisperse limit is recovered 
by decreasing the polydispersity. In a pressure-temperature representation 
this limit is highly singular: for example the slope of the perimeter of the 
coexistence region at the critical point will converge, as the monodisperse 
limit is taken, to a negative value -- even though the coexistence line (onto 
which the region itself collapses in this limit) has positive slope everywhere! 
This behavior is easily understood from Figure 2, and the Landau expansion 
seems to have no difficulty dealing with it.

The perturbation series in density that underlies the Landau expansion 
contrasts with, and is complementary to, a previous perturbative approach 
that was developed for the limit of narrow polydisperse distributions \cite{RML}. The latter approach always breaks down close enough to the 
critical point (even within mean field theory, to which, unlike our work, it 
is not limited in principle).

Finally, we note that although our Landau expansion was 
derived within the specific context of Eq.\ (\ref{Energy0}), the generic features 
of the phase diagram that it reproduces are seen rather generally in 
experiments and other theoretical models. We have not 
attempted explicit calculation for the case where more than one moment is 
present in the excess free energy, but since the topology of the phase 
diagram (near the fluid-fluid critical point) is already captured by 
Eq.\ (\ref{Energy0}), it is possible that any such refinements will not 
alter the basic structure of these results. A firm conclusion on this 
point is, however, left to future work. 

{\it Acknowledgements} We are grateful to Yuri Mart\'{\i}nez-Rat\'{o}n,
Jos\'{e} Cuesta, Patrick Warren and Peter Sollich for discussions. 
Work funded by EPSRC Grant GR/M29696. C.R.\ also acknowledges support
by the Ministerio de Ciencia y Tecnolog\'{\i}a (Spain) under grant 
BFM2000-0004.

\begin{figure}[t]
\centerline{\epsfig{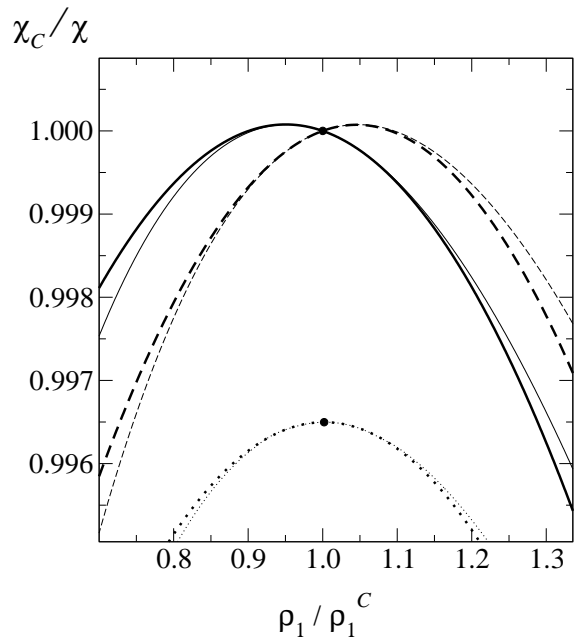}}
\vspace*{.75cm}  
\caption{Cloud (continuous) and shadow (dashed) curves of the
polymeric system described in the text. The monodisperse limit, 
$\Delta N\!\rightarrow\!0$, for which cloud and shadow curves 
merge into a single curve (the binodal), is also plotted as a 
dotted line. The thin lines represent the exact numerical solution 
of the model, while the thick lines represent the Landau expansion, 
which shows exact first and second order derivatives at the critical 
point. In the monodisperse case, the critical point rests at the
top of the curve.}
\label{fig1}
\end{figure}

\begin{figure}[t]
\centerline{\epsfig{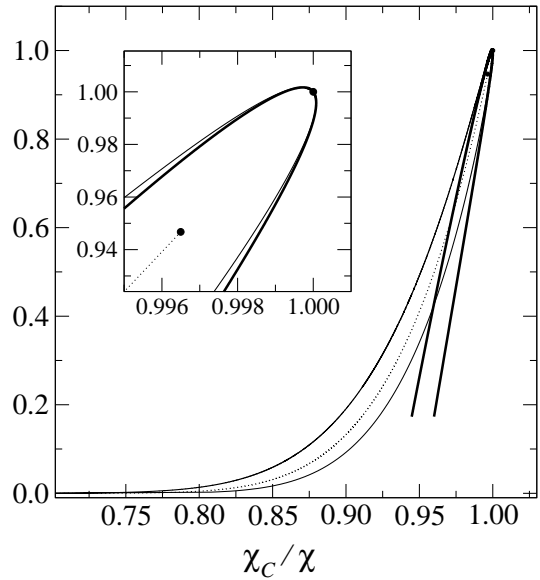}}
\vspace*{.75cm}  
\caption{$p\!-\!T$ phase diagram of the polymeric system 
described in the text. The thin line represents the exact 
numerical solution of the model, while the thick lines
represent the Landau expansion. This curve shows exact first and second
order derivatives at the critical point. The monodisperse limit
is also shown (dotted line). Inset: Magnified view of the
critical point.}
\label{fig2}
\end{figure}

\begin{table}[t]
\begin{tabular}{|c|l|l|}
Quantity & Landau Coefficients & Moments \\ \hline 
$\alpha_{\scriptscriptstyle CL}$ & $A_1$,$A_2$ & $n_1$,$n_2$,$n_3$ \\ \hline 
$\alpha_{\scriptscriptstyle SH}$ & $A_1$,$A_2$,$B_1$,$B_2$,$C_0$ &
$n_1$,$n_2$,$n_3$,$n_4$ \\ \hline
$\beta_{\scriptscriptstyle CL}$ & $A_{1-4}$, 
$B_1$,$B_2$,$C_0$  &
$n_1$,$n_2$,$n_3$,$n_4$ \\  \hline
$\beta_{\scriptscriptstyle SH}$ & $A_{1-4}$, $B_{1-4}$,
$C_0$,$C_1$,$C_2$,$D_0$  &
$n_1$,$n_2$,$n_3$,$n_4$,$n_5$\\ 
\end{tabular}
\caption{Landau coefficients and moments needed to calculate slope and curvature parameters of the cloud and shadow.}
\end{table}


\begin{references}
\bibitem{Ideal} Note that, with this definition, the {\it ideal} part of 
the free energy does not include the entropy of the solvent.
\bibitem{Vanish} Any of those coefficients can be written as
\begin{eqnarray}
\label{Vanish0}
\frac{\partial^{\,i}}{\partial{P_0}^i}\, 
\frac{\partial^{\,j}}{\partial{\rho_1}^j}\,
\left(\frac{1}{\rho_2}\right)
\end{eqnarray}
for appropriate $\;i\ge 1$ and $j\ge 0$ ($j\!=\!0$ means no derivative).
The first $j$ derivatives produce a certain function of moments 
(via Eq.\ (\ref{PropRho1})), while the following $i$ derivatives
produce terms vanishing in the monodisperse limit (via Eq.\ (\ref{PropRho2})).
\bibitem{sign} Since, for any (not infinitely narrow) distribution, 
$n_1n_3-n_2^2=\frac{1}{2}\int\!d\sigma d\sigma'\,
n(\sigma)n(\sigma')w_1(\sigma)w_1(\sigma')(w_1(\sigma)-w_1(\sigma'))^2$,
is necessarily positive if $w_1(\sigma)$ is.
\end{references}
\end{document}